\documentclass[twocolumn]{aa} 

\usepackage{natbib}
\usepackage{amsmath}
\usepackage{graphicx}
\usepackage{txfonts}
\usepackage{textcomp}
\usepackage{booktabs}
\usepackage{calligra}
\usepackage{threeparttable}
\usepackage[T1]{fontenc}
\makeatletter
\def\@biblabel#1{} 
\makeatother

\newcommand{\mjup}{M$_{\rm J}$\,}
\newcommand{\rjup}{R$_{\rm J}$\,}
\newcommand{\msun}{M$_\odot$\,}
\newcommand{\rsun}{R$_\odot$\,}
\newcommand{\mstar}{M$_\star$}
\newcommand{\lstar}{L$_\star$}
\newcommand{\rstar}{R$_\star$}
\newcommand{\mplan}{M$_p$}
\newcommand{\rplan}{R$_p$}
\newcommand{\ms}{m\,s$^{-1}$\,}
\newcommand{\teff}{T$_{{\rm eff}}$\,}
\newcommand{\logg}{log\,$g$}
\renewcommand{\cite}{\citealp}

\begin{document}

   \title{A hot Saturn on an eccentric orbit around the giant star EPIC\,228754001}

  \titlerunning{}

   \author{M.~I. Jones \inst{1}
           \and R. Brahm \inst{2,3}
           \and N. Espinoza \inst{3,2}
           \and A. Jord\'an \inst{3,2,4}
           \and F. Rojas \inst{2}
           \and M. Rabus \inst{4,5}
           \and H. Drass \inst{5}
           \and A. Zapata \inst{5}
           \and M. G. Soto \inst{6}
           \and J. S. Jenkins \inst{6} 
           \and M. Vu\v{c}kovi\'c \inst{7}
           \and S. Ciceri \inst{8}
           \and P. Sarkis \inst{4}
           }
         \institute{European Southern Observatory, Casilla 19001, Santiago, Chile \\\email{mjones@eso.org}
         \and Millennium Institute of Astrophysics, Santiago, Chile
         \and Instituto de Astrof\'isica, Facultad de F\'isica, Pontificia Universidad Cat\'olica de Chile, Av. Vicu\~{n}a Mackenna 4860, 7820436 Macul, Santiago, Chile
         \and Max-Planck-Institut f\"ur Astronomie, K\"onigstuhl 17, 69117 Heidelberg, Germany 
         \and Center of Astro-Engineering UC, Pontificia Universidad
         Cat\'olica de Chile, Av. Vicu\~{n}a Mackenna 4860, 7820436 Macul, Santiago, Chile 
         \and Departamento de Astronom\'ia, Universidad de Chile, Camino El Observatorio 1515, Las Condes, Santiago, Chile
         \and Instituto de F\'isica y Astronom\'ia, Universidad de Vapara\'iso, Casilla 5030, Valpara\'iso, Chile
         \and Department of Astronomy, Stockholm University, 11419, Stockholm, Sweden
         }

   \date{}


  \abstract {Although the majority of radial velocity detected planets have been found orbiting solar-type stars, a fraction of them have been discovered around giant stars. These planetary systems have revealed different orbital properties when compared to solar-type stars companions.
In particular, radial velocity surveys have shown that there is a lack of giant planets in close-in orbits around giant stars, in contrast to the known population of {\it hot-Jupiters} orbiting solar-type stars. The reason of this distinctive feature in the semimajor-axis distribution has been theorized to be the result of the stellar evolution and/or due to the effect of a different formation/evolution scenario for planets around intermediate-mass stars. However, in the past few years, a handful of transiting short-period planets (P$\lesssim$ 10 days) have been found around giant stars, 
thanks to the high precision photometric data obtained initially by the {\it Kepler} mission, and later by its two-wheels extension {\it K2}. These new discoveries, have allowed us for the first time to study the orbital properties and physical parameters of these intriguing and elusive sub-stellar companions.\newline
In this paper we report on an independent discovery of a transiting planet in field 10 of the {\it K2} mission, also reported recently by Grunblatt et~al. (\cite{GRU17}). The host star has recently evolved to the giant phase, and has the following atmospheric parameters:  \teff=4878 $\pm$ 70 $K$, \logg = 3.289 $\pm$ 0.004 and [Fe/H] = -0.11 $\pm$ 0.05 dex. The main orbital parameters of EPIC\,228754001\,$b$, obtained with all the available data for the system, are the following: $P$ = 9.1708 $\pm$ 0.0025 $d$, $e$ = 0.290 $\pm$ 0.049, \mplan = 0.495 $\pm$ 0.007 \mjup \,and \rplan = 1.089 $\pm$ 0.006 \rjup. This is the fifth known planet orbiting any giant star with $a < 0.1$, and the most eccentric one among them, making EPIC\,228754001\,$b$ a very interesting object.
}


   \maketitle

\section{Introduction}

To date, more than 3000 planetary companions have been discovered\footnote{As of June 2017; source: http://exoplanets.org}
orbiting stars other than 
the Sun, and this number is rapidly evolving as more and more new planets are routinely detected by different groups.
Strictly speaking, the first extrasolar planetary system was found around a stellar remnant, namely the pulsar PSR 1257+12
(Wolszczan \& Frail \cite{WOL92}). However, a couple of years later, Mayor \& Queloz (\cite{MAY95}) announced the detection
of a periodic signal in the radial velocity (RV) observations of the solar-type star 51 Pegasi.
The signal was caused by the presence of a giant planet in a 4-day orbit, confirming the existence of extrasolar planets. 
This discovery marked the beginning of the exoplanet observational area, which is currently living a golden age. 
\newline \indent
Afterward, new RV measurements allowed the detection of several {\it hot-Jupiters} and a large fraction 
of eccentric planetary companions (e.g. Marcy et al. \cite{MAR05}), which completely changed our knowledge of planetary
formation and evolution, that was mainly restricted to the study of the Solar System. These discoveries severely challenged the
planet formation theories, bringing back to life the importance of dynamical processes like planet migration
(Papaloizou \& Lin \cite{PAP84}) and eccentricity excitation via planet-star (e.g Kozai \cite{KOZ62}; Lidov \cite{LID62})
and planet-planet interactions (e.g. Rasio \& Ford \cite{RAS96}; 
Weidenschilling \& Marzari \cite{WEI96}; Lin \& Ida \cite{LIN97}). 
Moreover, as soon as instruments like HARPS (Mayor et al. \cite{MAY03}) capable of reaching $\lesssim$ 1\,\ms
precision were developed, a large population of small rocky planets was unveiled,
whose RV signals were hidden behind the instrumental noise and the stellar jitter (e.g., Mayor et~al. 2009). \newline \indent
Similarly, pioneering studies aimed at detecting transiting planets from ground-based photometric data (Charbonneau et al. \cite{CHA00}, Henry et~al.\ 2000),
led to surveys that discovered of a multitude of short-period giant planets (e.g., Bakos et~al. 2004, Pollacco et~al. 2006), providing direct
information of their physical properties, such as the density, radius and atmospheric composition (Seager \& Deming 2010, Crossfield 2015). Moreover, when combined with RV data, the planet mass can be directly inferred as well as the spin-orbit angle
from the Rossiter-McLaughlin effect (e.g., Queloz et al. \cite{QUE00}; Brown et al. \cite{BRO12}). 
However, only the advent of dedicated space-based missions like {\it CoRoT} (Baglin et al. \cite{BAG06}) and
{\it Kepler} (Borucki et al. \cite{BOR10}), has allowed us to efficiently detect transiting rocky planets, whose
transit depth are as small as $\sim$ 100 ppm. Similarly, space-based observations have more recently permitted
the detection of transiting planets orbiting around giant stars, which is incredibly challenging from the Earth, due to the small transit depth and long duration of the transit. However, these systems are
of great importance for several reasons. First, by studying the planet radius as a function of the stellar irradiation
(see Demory \& Seager \cite{DEM11}), it is possible to discriminate between the direct inflation scenario due the increasing
stellar irradiation as the host star evolve
through the giant phase (Grunblatt et al. \cite{GRU16}) or due to delayed thermal contraction (Lopez \& Fortney \cite{LOP16}).
Second, RV surveys have found an intriguing lack of short-period (P $<$ 10 days) giant planets around evolved stars
(e.g. Johnson et al. \cite{JOH07}; D\"ollinger et al. \cite{DOL11}; Jones et al. \cite{JON14}), in direct contrast to what is
observed in solar-type host stars.
In fact, despite the fact that over a thousand of such post main-sequence (MS) stars have been targeted by different groups
(Frink et al. \cite{FRI01}; Setiawan et al. \cite{SET03}; Hatzes et al. \cite{HAT05}; Sato et al. \cite{SAT05};
Niedzielski et al.  \cite{NIE09}; Jones et al. \cite{JON11}; Wittenmyer et al. \cite{WIT11}), only one short-period planet has
been detected by means of RV measurements (Johnson et al. \cite{JOH10}). Therefore, the detection of new
giant transiting planets around giant stars provides us with valuable information about the properties of these
elusive substellar companions. \newline \indent
In this paper we present the discovery of a Saturn-mass planet in a short-period  and eccentric orbit around the evolved star
EPIC 228754001. The transit signal was detected from {\it K2} (Howell et al. \cite{HOW14}) photometric data taken in Campaign 10, as part of a Chilean-based effort aimed at the detection and characterization of transiting exoplanetary systems (see Brahm et al. \cite{BRA16}, Espinoza et al. \cite{N16}, Espinoza et al. \cite{N16hj}). Additionally, we performed a spectroscopic follow-up using HARPS and FEROS. From these
datasets we computed precision RVs, which confirm the transit signal of the companion. From the combined transit and photometric data we obtained the following planet parameters: $P$ = 9.171 $^{+0.002}_{-0.003}$ $d$, $e$ = 0.29 $^{+0.05}_{-0.05}$, \mplan = 0.495 $^{+0.006}_{-0.007}$ \mjup and \rplan = 1.089$^{+0.008}_{-0.008}$. \newline \indent
The paper is structured as follows. In section 2 we present the photometric analysis and RV measurements. In section 3 we
describe the host star properties, including the asteroseismic analysis,  and the global modeling of the photometric and RV data.
Finally, the discussion and summary are presented in section 4.

\section{Observations}

\subsection{K2 Photometry}
\begin{figure*}
\centering
\includegraphics[width=2\columnwidth]{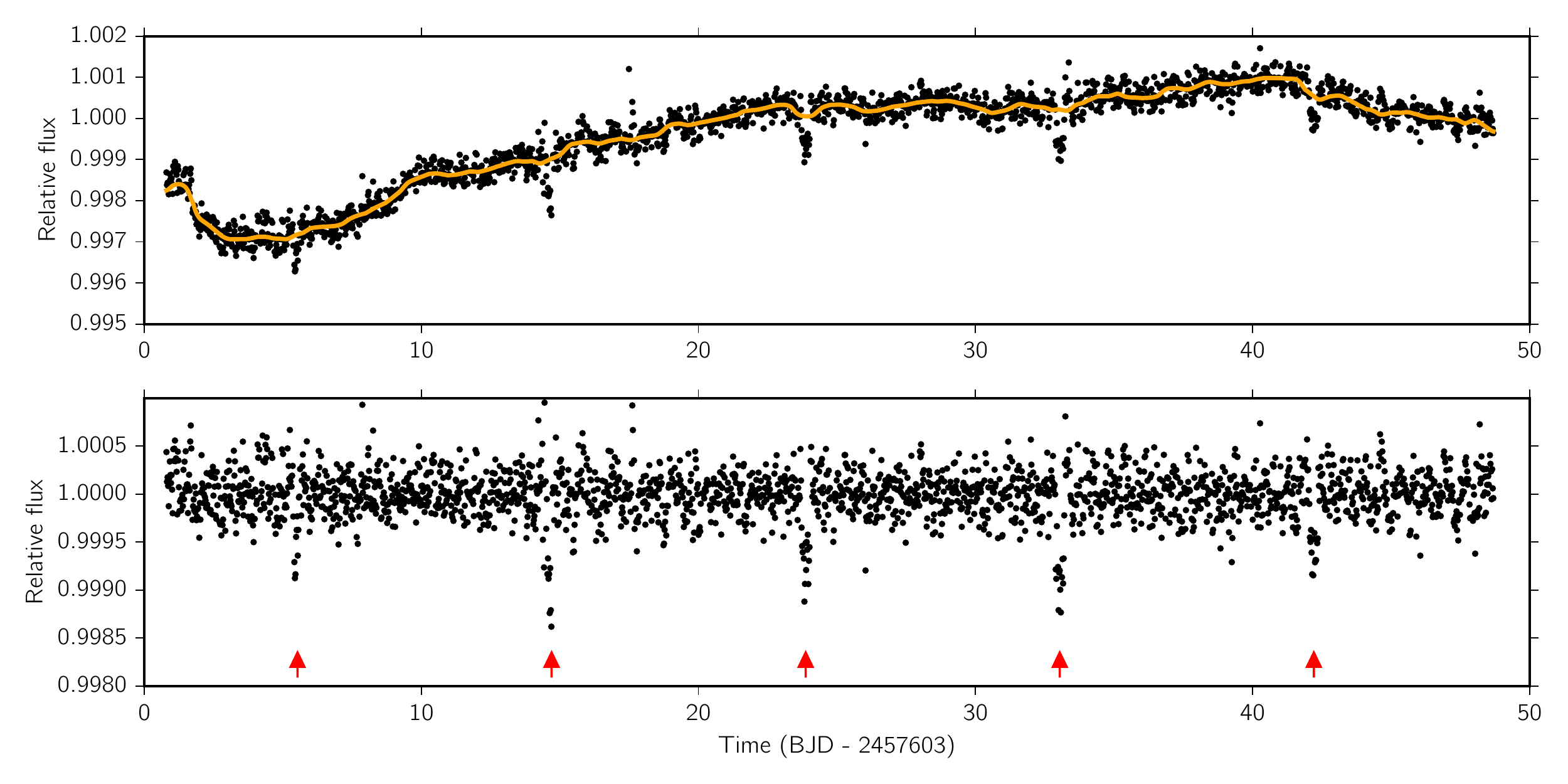}
\caption{Top panel -- Lightcurve of EPIC 228754001 (black points) along with the smoothed median filter used to model its long-term trend (orange solid line). Bottom panel -- Lightcurve detrended by the long-term trend. The transits detected by our BLS analysis are indicated with red arrows.
\label{k2photometry}}
\end{figure*}

Photometry for the star EPIC 228754001 was obtained in the long-cadence mode with the {\it Kepler} spacecraft during Campaign 10 of the repurposed {\it K2} mission. As described in previous works (Espinoza et al. \cite{N16}; Brahm et al. \cite{BRA16}), photometry was obtained for all stars in the field using our own implementation of the EVEREST algorithm (Luger et al. \cite{LUG16}) as soon as the data was available at MAST. A box least-squares (BLS) algorithm was used in order to search for planetary signals in each of the lightcurves, after these were detrended by any long-term trends using a median filter smoothed with a Gaussian filter. Our algorithm detected transit-like features with depths of $\sim 1000$ ppm and a period of $\sim 9$ days in the lightcurve of EPIC 228754001, and thus it entered in our list of transiting planet candidates and was selected for further spectroscopic follow-up in order to confirm or reject its possible planetary nature. 
In the reminder of this paper, we prefer to use the lightcurves of Vanderburg \& Johnson (\cite{VAN14}) due to the fact that these attain better precisions than our lightcurves obtained using our implementation of the EVEREST algorithm. Figure \ref{k2photometry} shows this photometry, where the transits can be spotted by eye. The first two transits of the planet candidate observable in this lightcurve showed strong systematics, and we decided to not include those in our analysis. 

\subsection{Precision Radial Velocities}
To confirm the planetary nature of the transiting candidate EPIC\,228754001 identified from the K2 photometry, we
obtained high resolution spectra using two different environment stabilized instruments. These observations were
used for: i) perform a fast spectral classification of the star for determining the expected size of the planet and the expected amplitude of the radial velocity signal, ii) to identify if the observed spectrum is composed of more that one stellar spectrum which could imply that the photometric signal is produced by a blended eclipsing binary, and iii) measure precise radial velocities for ruling out the presence of stellar companions and determine the orbital parameters of the planetary system.

We obtained the first spectra of EPIC\,228754001 using the HARPS spectrograph mounted to the ESO 3.6m telescope at the ESO La Silla Observatory on April 23, 2017. From this first spectrum we were able to identify that the star was an early K-type (\teff = 4900 $\pm$ 200 K) star with relatively low surface gravity (\logg = 3.25 $\pm$ 0.3 dex) consistent with a red giant star, which implied that the transits could have been produced by a giant planet.
We obtained another seven HARPS spectra for EPIC\,228754001
between April and May of 2017, to measure the radial velocity variations. We used exposure times of 1200 seconds to achieve a typical signal-to-noise ratio of $\sim$30, which produces photon noise dominated errors in RV of the order of 7 m s$^{-1}$.
Given that the nightly instrumental velocity drift of this spectrograph is significantly smaller than the expected errors in RV, we did not use the simultaneous comparison fibre. The spectra were reduced and analyses with the CERES pipeline (Brahm et al. \cite{ceres}), which performs the optimal extraction and wavelength calibration of the spectra, before computing the corresponding RVs, bisector spans, and a rough spectral classification.

We obtained six additional spectra of EPIC228754001 between May and June of 2017, using the FEROS spectrograph (Kaufer et al. \cite{KAU99}) mounted on the 2.2m MPG telescope at the ESO La Silla Observatory. In this case, the amplitude of the instrumental drift during one night if of the order $\sim$200 m s$^{-1}$, and therefore we used the comparison fibre to trace the instrumental velocity drift by using a ThAr lamp. We used exposure times of 1200 seconds which delivered radial velocity errors of $\sim$7 m s$^{-1}$. The FEROS spectra were also reduced and analyzed using CERES.

During the final writing phase of this article, Grunblatt et al. (\cite{GRU17}) announced the discovery and characterization of this same target, and published radial velocities obtained with KECK/HIRES. For completeness, we also include that data in our analysis which as will be seen in the next section, allowed us to further refine the system parameters.

\begin{figure}[tbp]
\centering
\includegraphics[width=\columnwidth]{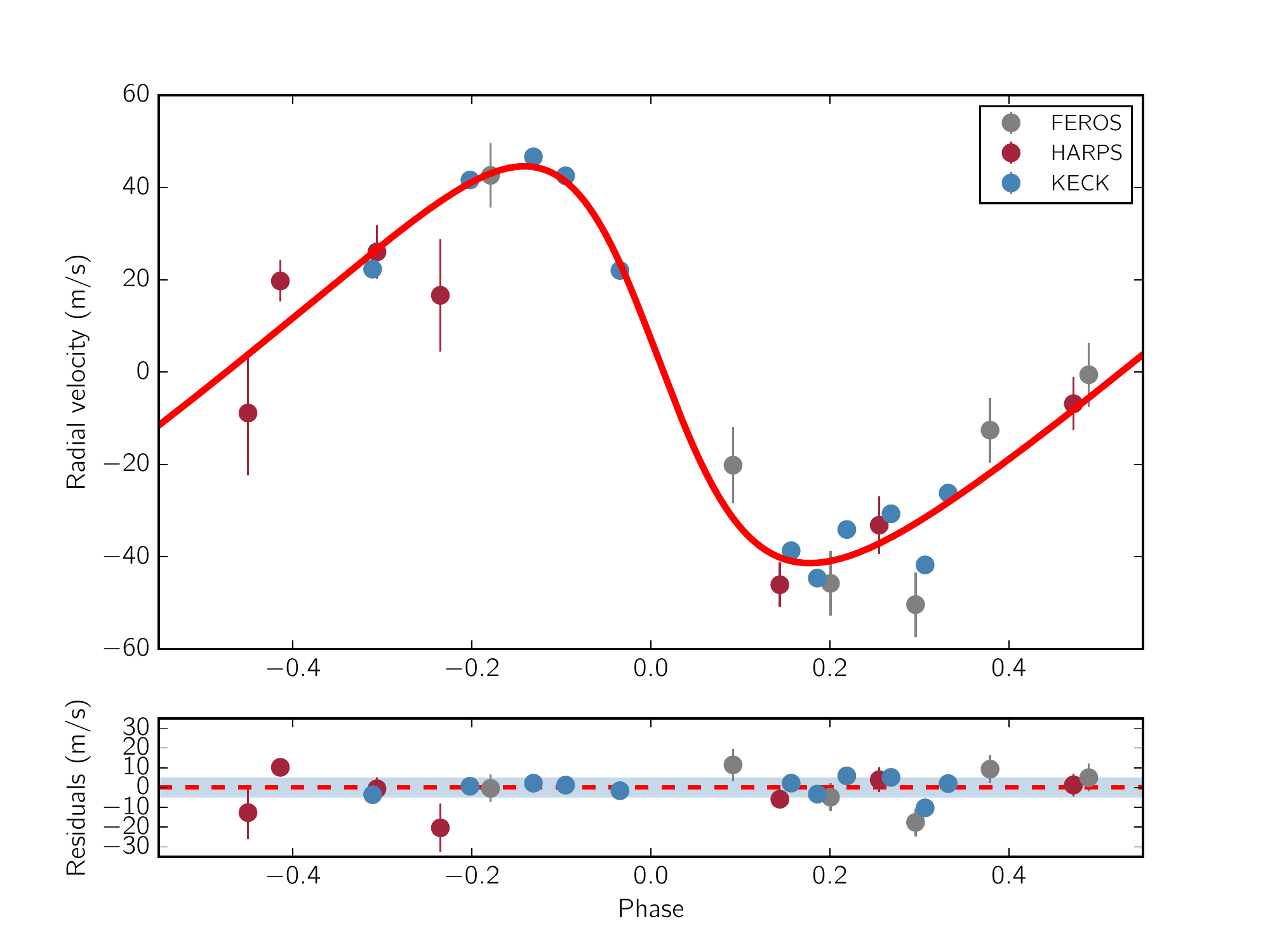}
\caption{Top panel -- Phased-folded radial velocities of EPIC\,228754001 obtained with HARPS (red dots), FEROS (gray dots) and KECK/HIRES (blue dots). The red line corresponds to the Keplerian model using the posterior parameters of the MCMC analysis of Section \ref{global}. Bottom panel -- Residuals for the observed radial velocities. The small blue band around zero has a width equal to the fitted jitter term of $5.1$ \ms.
\label{rvs}}
\end{figure}

\begin{figure}[!tbp]
  \centering
  \begin{minipage}[b]{0.4\textwidth}
    \includegraphics[width=\textwidth]{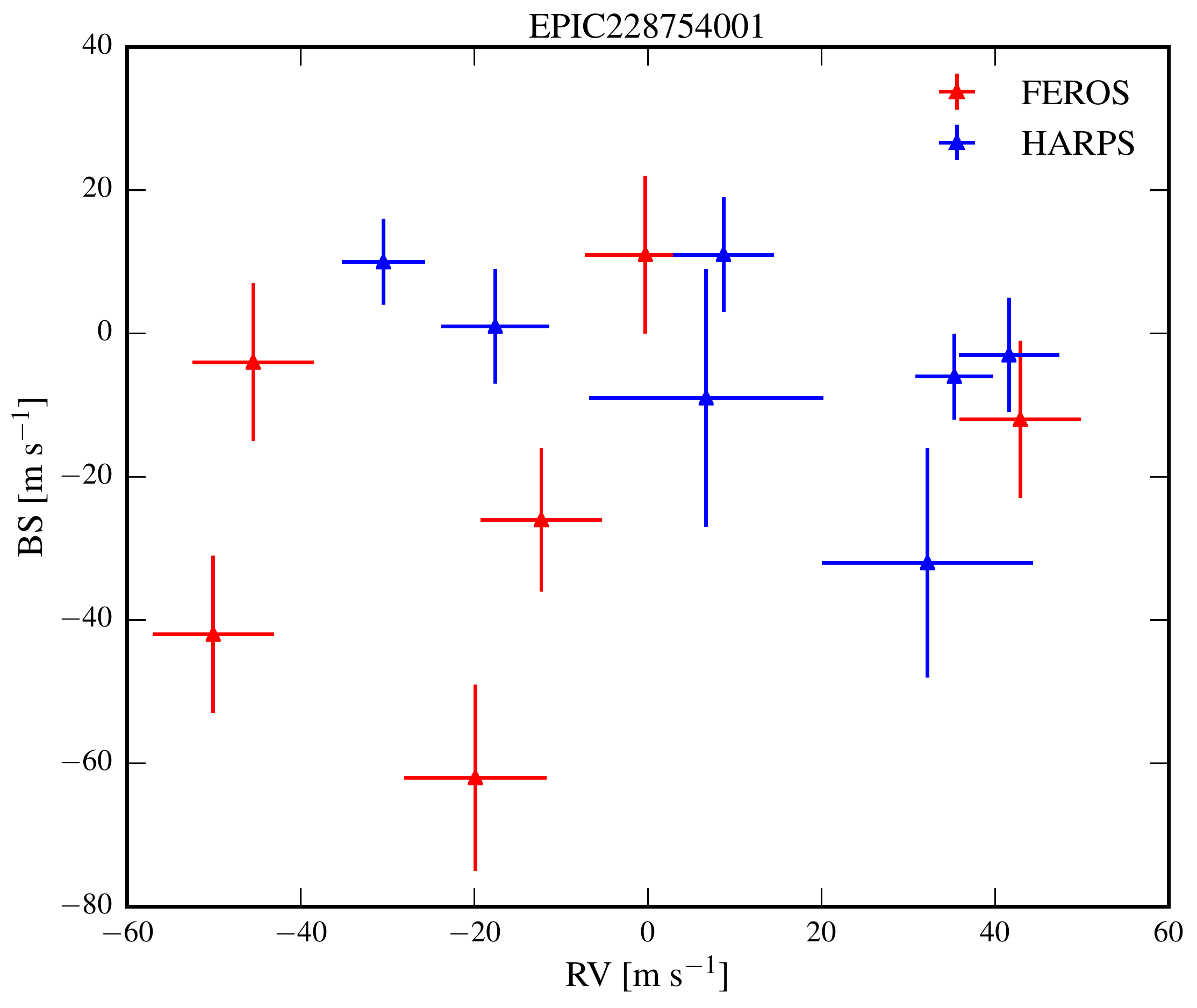}
  \end{minipage}
  \hfill
  \begin{minipage}[b]{0.4\textwidth}
    \includegraphics[width=\textwidth]{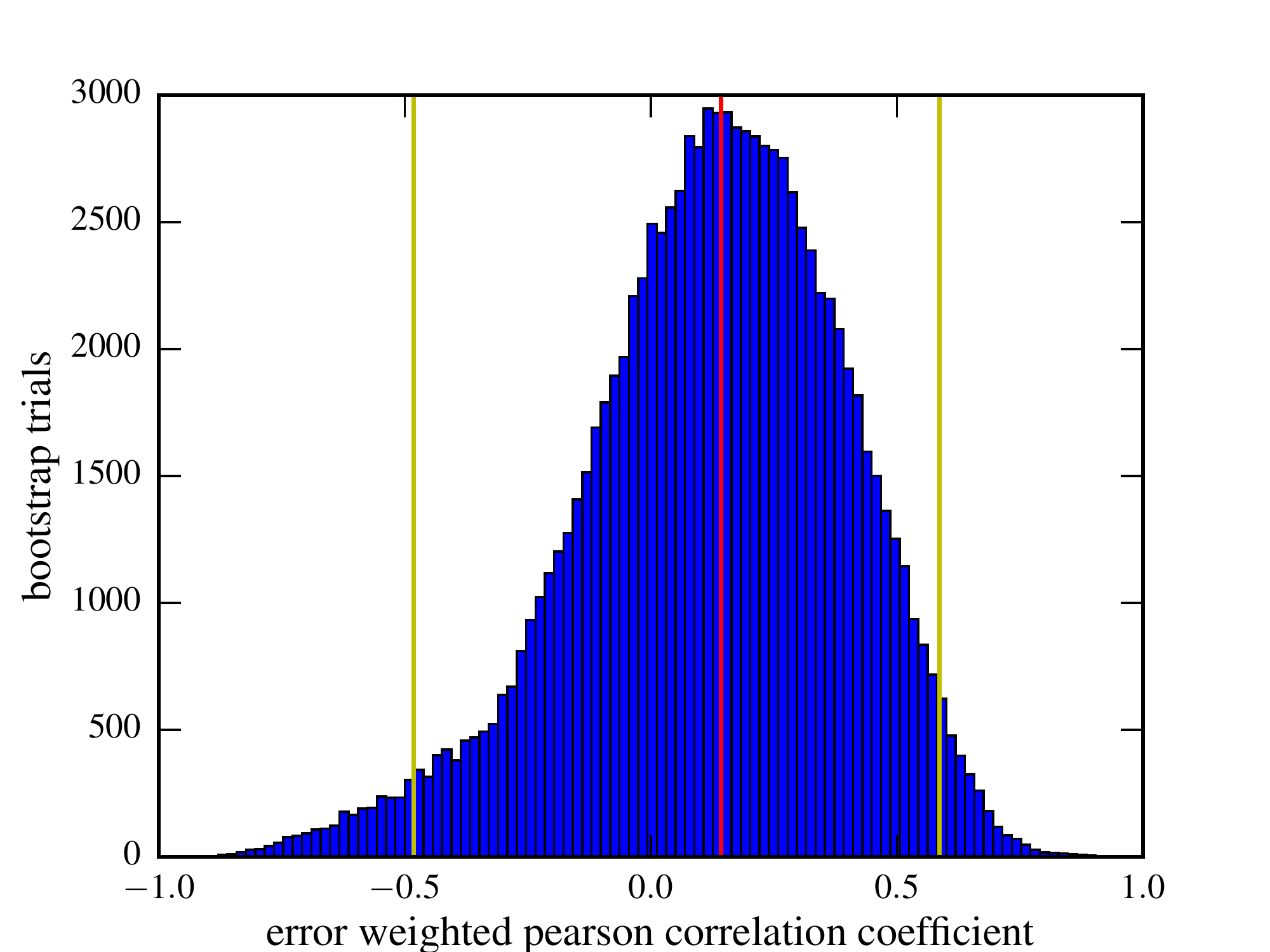}
  \end{minipage}
 \caption{Top -- observed radial velocities vs. bisector span measurements for HARPS (blue) and FEROS (red). Bottom -- histogram showing the bootstrap distribution of the weighted Pearson correlation coefficient between radial velocities and bisector span measurements. FEROS and HARPS velocity points are consistent with no correlation.\label{rvsbs}}

\end{figure}

As shown in Figure \ref{rvs}, the radial velocity variations measured by HARPS, FEROS and KECK/HIRES are consistent with a Keplerian orbit produced by a giant planet (K$\sim$40 m s$^{-1}$), and consistent with the photometric ephemeris of the {\it K2} light-curve. We also computed the degree of correlation between RVs and bisector span values for HARPS and FEROS (the analysis of the KECK/HIRES data can be seen in Grunblatt et al. \cite{GRU17}) in order to rule out the possibility that the observed RV variations are due to a blended scenario (Santerne et al. \cite{SAN15}). We used a bootstrap algorithm to determine the distribution of the error weighted Pearson correlation coefficient, finding that the data is consistent with no correlation at the 95\% confidence interval, as shown in Figure \ref{rvsbs}.

\section{Analysis}

\subsection{Atmospheric parameters}
To determine the atmospheric parameters of EPIC 228754001, we used the Zonal Atmospheric Stellar Parameters Estimator (ZASPE;
Brahm et al. \cite{BRA17}) code. Briefly, ZASPE matches the observed stellar spectrum with a set of synthetic spectra generated from the ATLAS9 model atmospheres (Kurucz \cite{kurucz}). This procedure is performed via
a global $\chi^2$ minimization, in a set of selected spectral regions, that are
highly sensitive to small changes in \teff, \logg\, and [Fe/H].
In addition reliable errors in the parameters are obtained by
considering the degree of systematic mismatch present between the observed spectrum and the optimal synthetic one.
In this specific case, we run ZASPE with a HARPS high S/N spectrum, which was built by combining eight individual spectra taken at different epochs, after correcting by their relative doppler shift. The results are summarized in Table \ref{atm_par}.
In addition, we obtained the atmospheric parameters by matching the curve of growth, using the equivalent width of a set of 
carefully selected Fe\,{\sc i} and Fe\,{\sc ii} lines (Jones et al. \cite{JON11}), and by imposing excitation and ionization
equilibrium. For this purpose we used the Spectroscopic Parameters and atmosphEric ChemIstriEs of Stars (SPECIES; Soto et al. in preparation) code, which implements an automated version of MOOG\footnote{http://www.as.utexas.edu/$\sim$chris/moog.html} (Sneden \cite{SNE73}) to iterate through the atmospheric parameters until the equilibrium condition is reached. 
The uncertainties in the results are derived by considering the contribution from the uncertainty in the excitation and ionization equilibrium, and from the correlations among the parameters.
These results are also listed in Table \ref{atm_par}. As can be seen, we obtained very good agreement between the results from ZASPE and SPECIES.

\subsection{Planet scenario validation}
To validate the planetary nature of EPIC 228754001\,$b$, we ran the Validation of Exoplanet Signals using a Probabilistic Algorithm (VESPA; Morton \cite{MOR12}). Since we do not detect any radial velocity consistent with a non-blended eclipsing binary system (see Section 2.2), we set the likelihood of this event to zero in these calculations, modeling then the possibility that our planet candidate could be produced by either a bona-fide planet, a blended eclipsing binary system or a hierarchical triple system. Assuming an occurrence rate of giant planets around giant stars similar to that of hot-Jupiters around solar-type stars ($\sim 1\%$; Marcy et al. \cite{MAR05}; Wang et al. \cite{WAN15}), we find a false-positive probability (FPP) for our system of $0.01\%$. If we consider a lower occurrence rate for giant stars of $\sim$ 0.1\%, then we obtain a FPP of $0.09\%$. 
This validates our system as a genuine exoplanet system. 

\subsection{Asteroseismology \label{sec:astero}}

\begin{figure}[!tbp]
\centering
\includegraphics[width=\columnwidth]{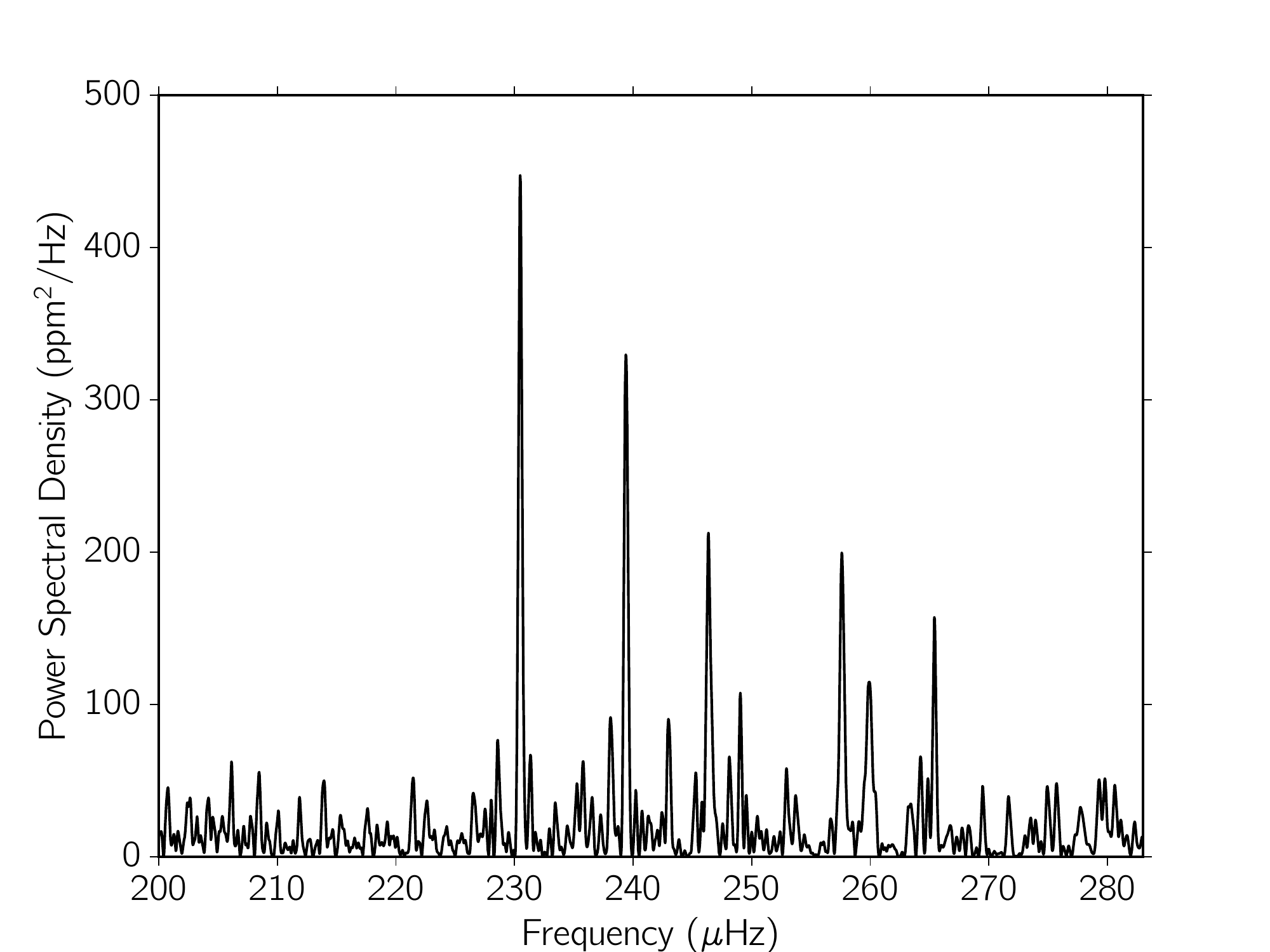}
\caption{Power Spectral Density (PSD) of the K2 lightcurve.
\label{psd-nofit-plot}}
\end{figure}

In Figure \ref{psd-nofit-plot} we show the Power Spectral Density (PSD) of the {\it K2} light curve shown in Figure \ref{k2photometry} with the transits removed. This shows a clear power excess with regularly spaced peaks at a frequency of $\approx 240 \mu$Hz, corresponding to a period of $\approx$ 70 minutes. Oscillations on that range are expected for a low luminosity giant (Bedding et al.~\cite{bedding:2010}), and an estimation of the frequency of maximum power, $\nu_{\rm max}$ and the large frequency separation, $\Delta \nu$, allow precise estimations of the stellar density and \logg. We estimate these parameters from the power spectrum using a method similar to that of Huber et al. (\cite{huber:2009}). To estimate $\nu_{\rm max}$ we smoothed the power spectrum with a Gaussian kernel with $\sigma=10\, \mu$Hz and take the maximum of the smoothed power spectrum as the estimate. The value of $\Delta \nu$ was estimated as the peak in the autocorrelation function as estimated from the power spectrum smoothed with a Gaussian kernel with $\sigma=1\, \mu$Hz. Of several peaks in the autocorrelation function, only the one consistent with the known relation between $\nu_{\rm max}$ and $\Delta \nu$ (Stello et al. \cite{stello:2009}) was analyzed.  Uncertainties on $\nu_{\rm max}$ and $\Delta \nu$ were estimated by recalculating these parameters on realizations of the power spectrum where correlated noise is added with properties estimated from the background away from the peaks. The values we obtain are $\nu_{\rm max} = 240 \pm 6 \,\mu$Hz and $\Delta \nu=17.6 \pm 0.3\,\mu$ Hz. We used equations (1) and (2) in Grunblatt et al. (\cite{GRU16}) in order to obtain estimates for the density $\rho= 0.0242 \pm 0.0008$ gr cm$^{-3}$ and surface gravity \logg=3.29 $\pm 0.02$. We note that given the similarity between EPIC\,228754001 and K2-97 it is appropriate to use the value of $f_{\Delta \nu}$ assumed in Equation 1 by Grunblatt et al. (\cite{GRU16}).

\subsection{Physical parameters}

To compute the physical parameters and evolutionary status of EPIC\,228754001 we used the Yonsei-Yale isochrones by searching for the stellar age and mass of the model that most
closely resembles the observed properties of the host star.
We used the \teff and [Fe/H] derived with ZASPE but given that the stellar \logg\, is not tightly constrained by spectroscopy, we used the $\rho_{\star}$ derived from our asteroseismic analysis as a luminosity indicator for the isochrones.

\begin{figure}[tbp]
\centering
\includegraphics[width=\columnwidth]{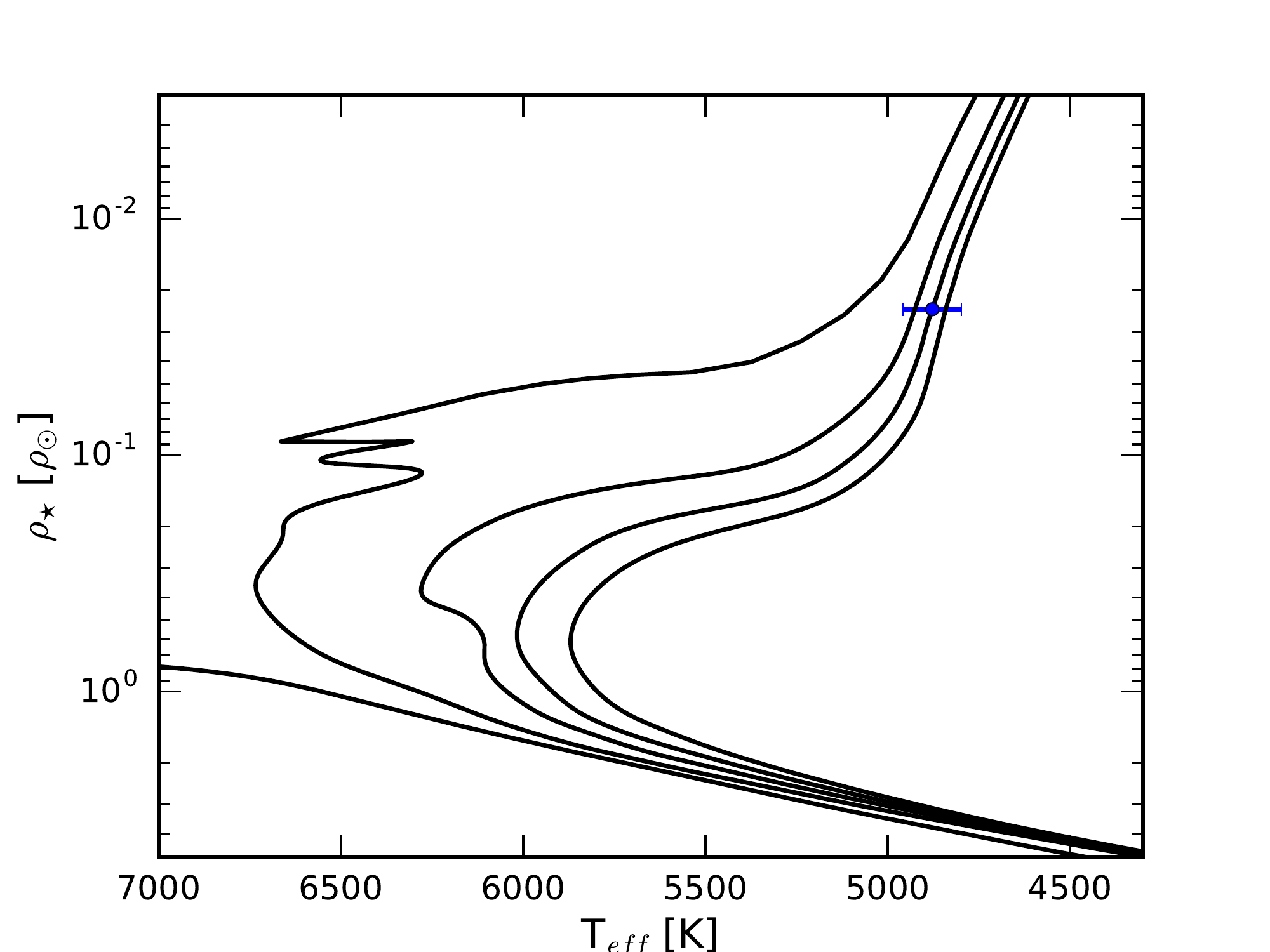}
\caption{Stellar density as a function of the stellar \teff for the Yonsei-Yale isochrones. From left to right the plotted isochrones correspond to ages of 0.1, 2, 4, 6, 8 Gyr. The blue dot corresponds to the assumed parameters of the host star.
\label{iso}}
\end{figure}

We run a Monte Carlo Markov Chain (MCMC) using the emcee Python package for exploring the parameter space. In this process we held fixed the [Fe/H] to the spectroscopic value, while the stellar age and mass were considered as free parameters. To compute the model \teff and $\rho_{\star}$ we interpolated the original isochrones in mass, age, and [Fe/H] using the algorithm provided with the isochrones.
Figure \ref{iso} displays some of the Yonsei-Yale isochrones  for the ZASPE determined metallicity for EPIC\,228754001 in the \teff -- $\rho_{\star}$ plane,
along with the corresponding assumed values for the
host star. 

\begin{table}
\centering
\caption{Stellar parameters of EPIC 228754001. \label{atm_par}}
\begin{tabular}{lrr}
\hline\hline
Parameter               &         Value       &       Method      \\
\hline
\vspace{-0.3cm} \\
\teff (K)               &  4878 $\pm$ 70      &       ZASPE       \\
                        &  4930 $\pm$ 36      &       SPECIES     \\
\logg \,(cm\,s$^{-2}$)  &  3.35 $\pm$ 0.15    &       ZASPE       \\
                        &  3.35 $\pm$ 0.89    &       SPECIES     \\
                        &  3.29 $\pm$ 0.02    &       Aster.      \\
{\rm [Fe/H]} (dex)      & -0.11 $\pm$ 0.05    &       ZASPE       \\
                        & -0.04 $\pm$ 0.08    &       SPECIES     \\
v\,sin$i$ (k\ms)        &  3.36 $\pm$ 0.25    &       ZASPE        \\
                        &  2.25 $\pm$ 0.74    &       SPECIES     \\
\mstar (\msun)          &  1.19 $\pm$ 0.04    &       Aster. + ZASPE +YY\\
                        &  1.16 $\pm$ 0.14    &       Aster.          \\
\rstar (\rsun)          &  4.11 $\pm$ 0.05    &       Aster. + ZASPE +YY\\
                        &  4.16 $\pm$ 0.20    &       Aster.          \\  
Age (Gyr)               &  5.5  $\pm$ 0.4      &    Aster. + ZASPE +YY  \\
\lstar (L$_\odot$)      &  8.78 $\pm$ 0.19  &    Aster. + ZASPE +YY\\
\rstar                  &  4.11 $\pm$ 0.05    &    Aster. + ZASPE +YY\\
\vspace{-0.3cm} \\\hline\hline
\end{tabular}
\end{table}

\subsection{Global Modeling}
\label{global}
The {\it K2} photometry along with the radial velocities measurements were fitted simultaneously with the \texttt{exonailer} algorithm\footnote{\url{https://github.com/nespinoza/exonailer}}, whose 
main characteristics are detailed in Espinoza et al. (\cite{N16}). In summary, we use a transit model using the \texttt{batman} package (Kreidberg \cite{KRE15}) and fit the transit lightcurve by resampling the transit model with the method of selective resampling described in Kipping (\cite{KIP13}). We follow Espinoza \& Jord\'an (\cite{ESP16}) and select the quadratic limb-darkening as the optimal law for this case, which provides the lowest mean-squared error in the planet-to-star radius ratio, which in this case is the most interesting parameter to retrieve. For the RVs, a different systemic velocity is fitted for each instrument, and a common jitter value is fitted simultaneously for every dataset, which is added in quadrature to the errorbars of each RV datapoint. We then use the \texttt{emcee} MCMC package to explore the parameter space and to obtain reliable estimates of the uncertainties of each parameter. 

As described in Section \ref{sec:astero}, the star clearly shows a correlated structure in the observed {\it K2} photometry, which is typical of red giant stars. In order to account for this structure, in our modeling we use the physically motivated Gaussian process (GP) model described in Foreman-Mackey et al. (\cite{FOR17}), which assumes that the power spectrum of the {\it K2} photometry can be described by two set of terms. The first one is a term which modulates the granulation ``background" of the process, which takes care of the power at lower frequencies and whose power spectral density (PSD) is given by
\begin{eqnarray*}
S(\omega) = \sqrt{\frac{2}{\pi}} \frac{S_g}{(\omega/\omega_g)^4 + 1}.
\end{eqnarray*}
The second set of terms account for the power excess at larger frequencies due to the asteroseismic oscillations described in Section \ref{sec:astero}. The idea is to model the ``peaks" observed at and around $\nu_\textnormal{max}$. These terms are given by
\begin{eqnarray*}
S_{j}(\omega) = \sqrt{\frac{2}{\pi}} \frac{S_{0,j}\omega_{0,j}^4}{(\omega^2 - \omega_{0,j}^2)^2 + \omega_{0,j}^2\omega^2/Q^2},
\end{eqnarray*}
where $\omega_{0,j} = 2\pi(\nu_\textnormal{max} +j\Delta \nu +\epsilon)$ and $S_{0,j} = (A/Q^2)\exp\left(-(j\Delta \nu + \epsilon)^2/(2W^2)\right)$. The number of terms $j$ is somewhat arbitrary, and define the number of peaks one would want to capture. Based on the number of peaks in the periodogram around $\nu_\textnormal{max}$ we choose a total of seven terms ($j=-3,-2,...,2,3$). Finally, we also fit for a photometric jitter term $\sigma_w$ which in the time domain allow us to estimate the underlying photon noise on top of this stochastic process. 

\begin{figure}[tbp]
\centering
\includegraphics[width=\columnwidth]{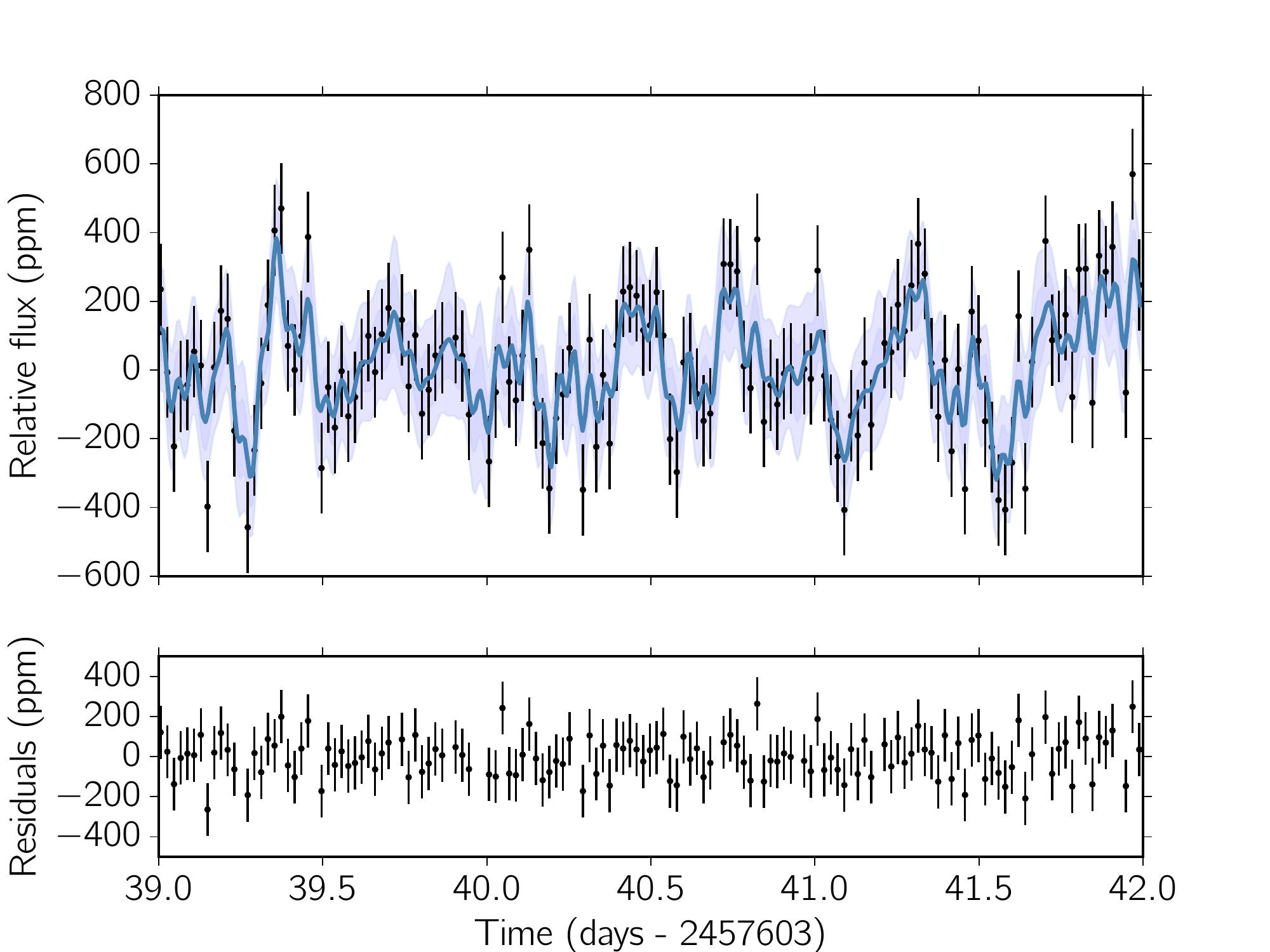}
\caption{Top panel -- Portion of the K2 photometry of the target star. The errorbars correspond to the fitted value of $\sigma_w$ using our noise model. The blue line with bands show the posterior prediction of our GP modeling at the given times and the 2-sigma credibility interval, respectively. Bottom panel -- Residuals between the fitted GP and the K2 photometry. No obvious structure is observed in the residuals.
\label{noise-plot}}
\end{figure}

\begin{figure}[tbp]
\centering
\includegraphics[width=\columnwidth]{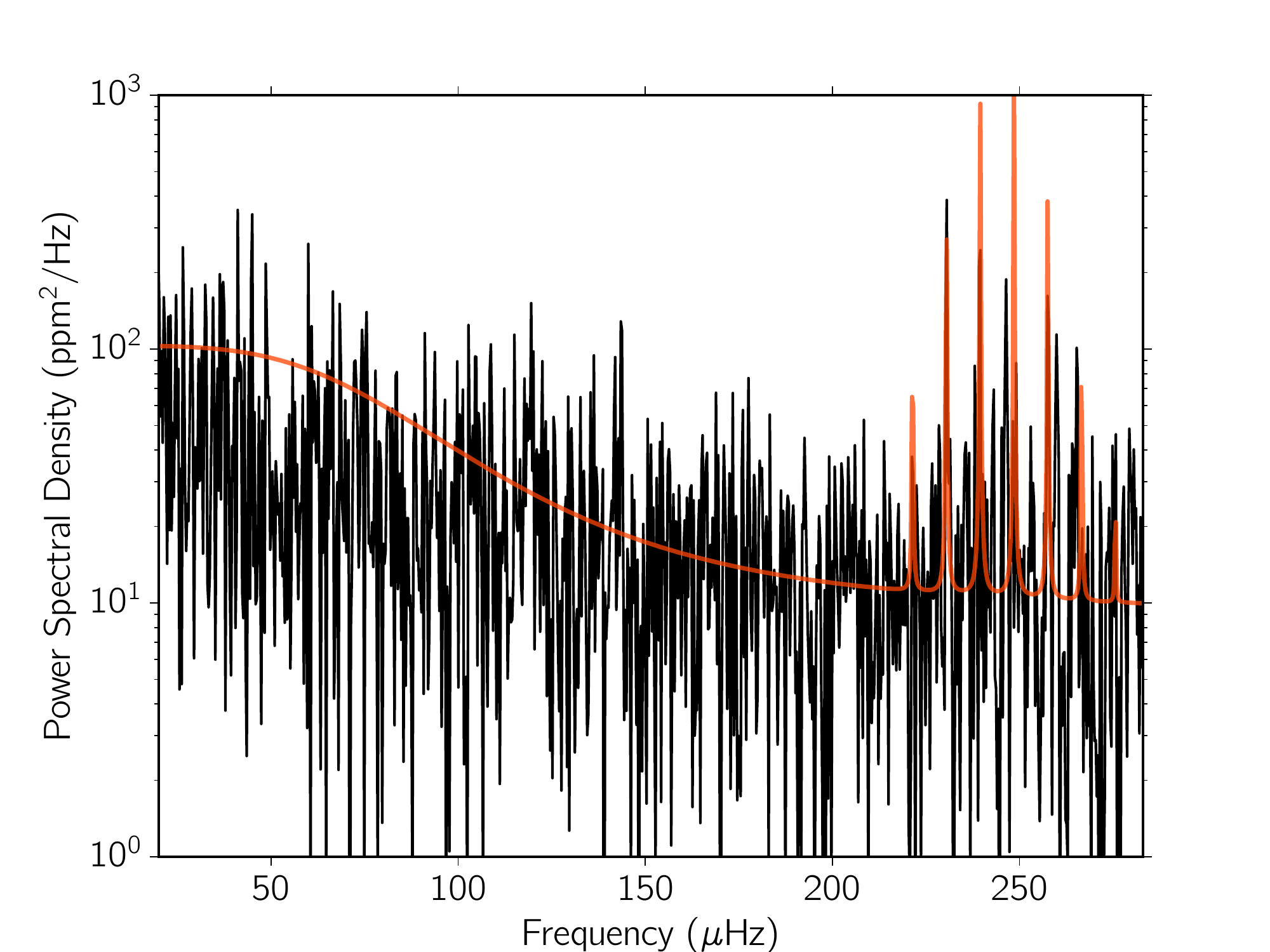}
\caption{Power Spectral Density (PSD) of the lightcurve shown in Figure \ref{noise-plot} (black) along with our noise model on top (red). Note how our ad-hoc model model captures the low-frequency ($\lesssim 200 \mu$Hz) and the high frequency components of the PSD. 
\label{psd-plot}}
\end{figure}

To model this correlated structure in the time-domain, we used the \texttt{celerite} package\footnote{\url{https://github.com/dfm/celerite}}, which we have implemented as part of \texttt{exonailer}. 
Given that the transits only occupy a small portion of the lightcurve, we decided to first analyze the lightcurve with the transits removed, estimate the parameters of the above defined noise model, and then use those parameters to account for the correlated structure in the transit and RV fitting. For this purpose, 500 walkers, with 1000 steps each are used, 500 of which are used as burn-in. The starting points of each of the parameters are based on a previous maximum-likelihood estimation of the parameters, and \texttt{emcee} is used to explore the parameter space in order to obtain parameter uncertainties. Table \ref{tab:noise-params} summarizes the results of this fit. Figure \ref{noise-plot} shows a 3-day portion of the lightcurve. It can be seen that our modeling captures the variability observed in the {\it K2} photometry in the time-domain. The PSD of the lightcurve is shown in Figure \ref{psd-plot} along with our model, which shows how the data and our model looks like in the frequency domain.

\begin{table}
\centering
 \caption{Noise parameters for the lightcurve of EPIC 228754001.}
 \label{tab:noise-params}
 \begin{threeparttable}
  \centering
  \begin{tabular}{ lcl }
   \hline
   \hline
     Parameter &  Prior & Posterior Value \\
   \hline
   \vspace{0.1cm}
~~~$\ln S_g$ \dotfill    & $\mathcal{U}(-15,15)$ & 6.60$^{+0.11}_{-0.11}$ \\
\vspace{0.1cm}
~~~$\ln \omega_g$ \dotfill    & $\mathcal{U}(-15,15)$ & 3.810$^{+ 0.099}_{-0.103}$ \\
\vspace{0.1cm}
~~~$\ln Q$ \dotfill    & $\mathcal{U}(-0.35,15)$ & 7.49$^{+0.96}_{-0.84}$ \\
\vspace{0.1cm}
~~~$\ln W$ \dotfill    & $\mathcal{U}(-4,4)$ & 2.53$^{+0.44}_{-0.31}$ \\
\vspace{0.1cm}
~~~$\ln A$ \dotfill    & $\mathcal{U}(-15,15)$ & 9.3$^{+1.3}_{-1.0}$ \\
\vspace{0.1cm}
~~~$\epsilon$ \dotfill    & $\mathcal{N}(0,11)$ & 6.2$^{+4.1}_{-3.7}$ \\
\vspace{0.1cm}
~~~$\ln \nu_\textnormal{max}$ \dotfill    & $\mathcal{U}(5.35,5.6)$ & 5.490$^{+0.015}_{-0.017}$ \\
\vspace{0.1cm}
~~~$\ln \Delta \nu$ \dotfill    & $\mathcal{U}(2,3)$ & 2.2009$^{+0.0035}_{-0.0052}$ \\
\vspace{0.1cm}
~~~$\sigma_w$ \dotfill    & $\mathcal{J}(10,1000)$ & 132.6$^{+7.2}_{-8.0}$ \\
\hline
   \end{tabular}
\textit{Notes}: All units except those of $\nu_\textnormal{max}$, $\Delta \nu$ and $\epsilon$, which are given in $\mu$Hz, are given in days and parts-per-million (ppm). $\mathcal{U}$ stands for a uniform distribution, $\mathcal{J}$ for a Jeffreys distribution and $\mathcal{N}$ for a normal distribution.
  \end{threeparttable}
 \end{table}

\begin{figure}
\centering
\includegraphics[width=\columnwidth]{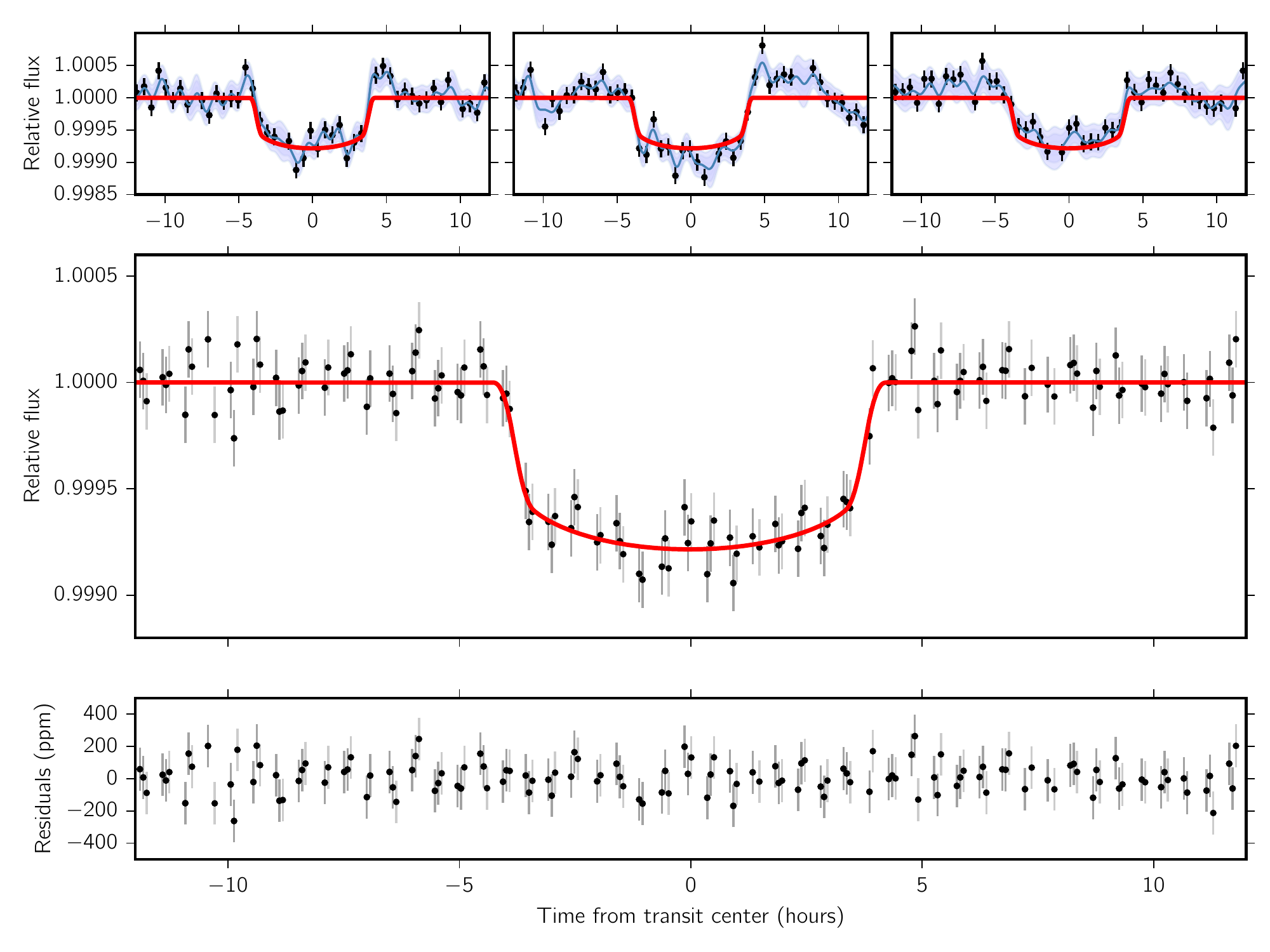}
\caption{Top panel -- Transits of EPIC228754001b (black points) along with the best-fit transit model (red solid line) and GP (blue line), along with the 3-sigma credibility interval for the posterior predictive GP regression (blue bands). Middle panel -- Phase-folded lightcurve with the GP model removed (black points) along with the best-fit transit lightcurve model (red). Bottom panel -- Residuals of the phase-folded lightcurve after removal of the best-fit transit and GP model.
\label{transit}}
\end{figure}

Having parametrized the noise in the lightcurve with the above mentioned modeling, we proceeded to fit the transit and radial velocities simultaneously using these noise parameters as inputs for the photometric modeling. We used the estimated stellar density in Section \ref{sec:astero} in order to put a prior on $a/R_*$ through the relation $a/R_* = \left(G\rho_*P^2/3\pi\right)^{1/3}$, where $P$ is the period of the orbit and which is very well constrained with our BLS analysis. We use 500 walkers with 1000 steps each in order to perform this simultaneous fit, where the first 500 steps are discarded as burn-in. Figures \ref{transit} and \ref{rvs} show the results of this simultaneous modeling, with the former presenting the transit lightcurve modeling and the latter presenting the modeling of our high precision RV measurements. Table \ref{glo_mod} summarizes the results of our global modeling .

Our modeling predicts a quite interesting eccentricity of $0.29\pm 0.05$. In order to test how this eccentricity is supported by the data, we repeated our modeling assuming a circular orbit and computed an estimate of the evidence of both models by using the Bayesian Information Criterion (BIC). The difference of the BIC values between the circular and eccentric models is $\Delta \textnormal{BIC} = 15.8$ in favor of the eccentric model. This implies that the eccentric model is $\exp (\Delta \textnormal{BIC}/2)\approx 2700$ times more likely than the circular model, which is strong evidence against this latter model. 

\begin{table}
\centering
\caption{Parameters obtained from the global modeling. \label{glo_mod}}
\begin{tabular}{lr}
\hline\hline
Parameter               &         Value       \\
\hline
\vspace{-0.3cm} \\
Light curve parameters  &  \\
\ \  P (days)               &  9.1708 $\pm$ 0.0025       \\
\ \  T$_c$ (BJD)            &  2457608.5289 $\pm$ 0.0087 \\
\ \  $a$/\rstar               &  4.76 $\pm$ 0.50    \\
\ \  R$_P$ / \rstar         &  0.0279 $\pm$ 0.0011    \\
\ \  $i$                      &  77.46$^{+0.47}_{-0.51}$  \\
\ \  c$_1$                  &  0.27$^{+0.24}_{-0.17}$   \\
\ \  c$_2$                  &  0.48$^{+0.35}_{-0.33}$   \\
 & \\
RV parameters \\
\ \  e                      &  0.290 $\pm$ 0.049    \\
\ \  $\omega$               &  $82.6^{+4.0}_{-4.2}$    \\
\ \  K (m s$^{-1}$)         &  43.0 $\pm$ 1.8    \\
\ \  $\mu_{HARPS}$ (km s$^{-1}$)  & 0.0038 $\pm$ 0.0019 \\
\ \  $\mu_{FEROS}$ (km s$^{-1}$)  & 10.3686$\pm$ 0.0037 \\
\ \  $\mu_{HIRES}$ (km s$^{-1}$)  & 10.3839$\pm$ 0.0030 \\
\ \  $\sigma_{RV}$  (km s$^{-1}$)  & 0.0052 $\pm$ 0.0010 \\
 & \\
Derived parameters \\
\ \ M$_P$ (M$_J$) & 0.495$_{-0.0063}^{+0.0068}$ \\
\ \ R$_P$ (R$_J$) & 1.089$_{-0.006}^{+0.006}$ \\
\ \ $a$ (AU)      & 0.0916$_{-0.0006}^{+0.0006}$ \\
\ \ T$_{eq}$ (K)  & 1586 $\pm$ 10  \\

\vspace{-0.3cm} \\\hline\hline
\end{tabular}
\end{table}

\section{Discussion}

\subsection{Short period planets around giant stars}

\begin{figure}[!tbp]
\centering
\includegraphics[width=\columnwidth]{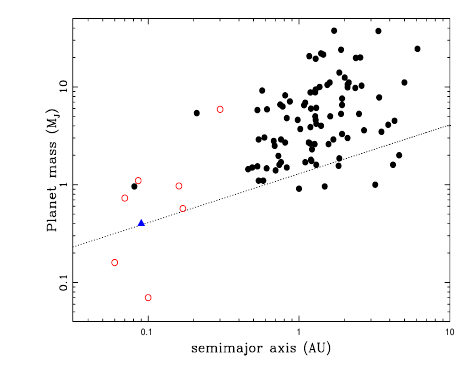}
\caption{Semimajor axis versus planetary mass of companions detected around giant stars. The black dots and open red circles correspond to planets detected via RV and transit method, respectively. The blue triangle corresponds to the position of EPIC\,228754001\,$b$.
\label{semiaxis_pmass}}
\end{figure}

One of the most intriguing results from RV surveys is the observed scarcity  of relatively close-in ($a \lesssim$ 0.5 AU) planets around post-MS stars. 
This observational trend has been attributed to be caused by the strong tidal torque exerted by the star, as its radius grow during the giant phase. As a result, planets are expected to lose orbital angular momentum, thus moving inward until they are evaporated in the stellar atmosphere (Livio \& Soker \cite{LIV83}; Sato et al. \cite{SAT08}; Villaver \& Livio \cite{VIL09}; Kunitomo, et al. \cite{KUN11}). 
On the other hand, the majority of the giant stars targeted by RV surveys are intermediate-mass stars (\mstar \,$\sim$ 1.5\,-\,3.0 \msun), thus they are the post-MS counterpart of A and early F main-sequence stars. Therefore, their companions should not be directly compared to those orbiting solar-type stars. Based on this analysis, known planets orbiting field giant stars are expected to be born in different conditions than those around low-mass stars. In particular, these planets are formed in more massive disks (since M$_d$ $\propto$ \mstar; Andrews et al. \cite{AND13}), from which they can efficiently accrete a significant amount of gas, becoming gas giants (e.g. Kennedy \& Kenyon \cite{KEN08}).
In addition, due to the higher gas accretion rate (Muzerolle et al. \cite{MUZ05}) and higher irradiation, these disks have shorter dissipation timescales (Currie \cite{CUR09}; Kennedy \& Kenyon \cite{KEN09}) and the snow line is located at a larger distance from the central star (Kennedy \& Kenyon \cite{KEN08}). 
As a consequence, these planets are most likely formed at larger orbital distance, and due to the shorter disk timescale inward migration is halted, thus reaching their final position at a relatively large distance from the parent star.
For comparison, Currie (\cite{CUR09}) predicted that only $\sim$ 1.5 \% of intermediate-mass stars host giant planets with $a \lesssim$ 0.5 AU, while $\gtrsim$ 7.5\% of them host at least one gas giants at $a \gtrsim$ 0.5 AU. Figure \ref{semiaxis_pmass} shows the mass versus the orbital distance of planets detected around giant stars (log$g$ $\lesssim$ 3.5), via RV measurements (black dots) and by the transit method (red open circles). We note that values of the RV detected systems correspond to the minimum planet mass (\mplan\,sin$i$). The dotted line represents a radial velocity semi-amplitude of $K$ = 30 \ms for a 1.5 \msun star, (corresponding to a 3\,$\sigma$ detection; e.g. Hekker et al. \cite{HEK06}). 
As can be seen, there is only one companion detected via RVs interior to 0.1 AU, and the rest of them reside at an orbital distance $a \gtrsim$ 0.4 AU.  
As discussed above, this observational result might be explained by the engulfment of the innermost planets as the parent star evolve off the MS, becoming a giant star. However, since a similar trend is observed in less evolved sub-giants, whose radii have not yet reached a value where tidal interactions are strong enough to affect the orbits of their companions, Johnson et al. (\cite{JOH07}) argued that this is probably explained by a different formation scenario between planets around low-mass stars and those formed in more massive disks. From Figure \ref{semiaxis_pmass} it is also evident that planets residing interior to $\sim$ 0.1 AU are significantly less massive (\mplan $\lesssim$ 1 \mjup) than those orbiting at a larger distance. In fact, two of these transiting 
planets are well below the 3\,$\sigma$ detection threshold, thus they are not detectable via radial velocities. A similar trend is also observed in MS stars (Zucker \& Mazeh \cite{ZUC02}), which might be caused by a decrease in the type II migration speed with increasing planetary mass, i.e., $da/dt$ $\propto$ M$_P^{-1}$ (Mordasini et al. \cite{MOR09}). This theoretical prediction naturally explains why the most massive planets are found at $a \gtrsim$ 0.4 A. 
On the other hand, the mass distribution of the parent stars of these two population of planets are different. In fact, while the mean stellar mass of the RV detected planets is 1.78 \msun, this value is only 1.38 \msun for the transiting systems and thus two distinct planet mass distributions are expected to be found. Moreover, a similar result is observed between the mass of planets orbiting sub-giant and giant stars (being planets around giant stars significantly more massive than those around sub-giants; see Jones et al. \cite{JON14}). In fact, the mean mass of the sub-giant parent stars is 1.5 \msun, significantly lower than giant host stars. These results provide further observational support of a different formation and migration scenario for planets at different host star mass. 

\subsection{Orbital evolution}

Short period planets are known to suffer from significant tidal interactions with their host stars. 
Observationally, this result is supported by the low orbital eccentricities of planetary companions with $a \lesssim$ 0.1 AU, when compared to farther-out planets (Marcy et al. \cite{MAR05}).
While the parent star is on the main-sequence, tides raised on the planet are thought to be the main responsible mechanism of the eccentricity damping, which at the same time produce a significant internal heating, and thus might explain the large observed radii of many transiting short period planets (Jackson et al. \cite{JAC08a}). 
However, after the host star evolve  to the giant phase, its radius rapidly increases, and tides raised on the stellar envelope become stronger, eventually dominating over tides within the planet (Schlaufman \& Winn \cite{SCH13}). We used equation (1) from Jackson et al. (\cite{JAC08b}) to compute the eccentricity damping timescale $\tau_e$ for 
EPIC,\,228754001\,$b$, including the contribution from both, tides raised in the planet and the star. We adopted the tidal quality factors of Q$_\star$ = 10$^{6.5}$ and Q$_p$ = 10$^{5.5}$, derived by Jackson et al. (\cite{JAC08b}).
Given the current radius of the host star, we obtained a circularization timescale of $\tau_e$ $\sim$ 3 Gyr. We note that at this point the effect of tides raised in the planet slightly dominate over tides in the star. Also, since $\tau_e$ is of the order of the age of the system, EPIC\,228754001\,$b$ has probably not suffered from significant eccentricity damping in the past.
However, since the host star is rapidly climbing the red giant branch (RGB), its radius is growing in timescales much shorter than $\tau_e$. 
For comparison, if we recompute $\tau_e$ when the star has reached about $\sim$ 8 \rsun (in $\sim$ 150 Myr from now), then the effect of the tides raised in the star completely dominate over tides in the planet, and we obtained a much shorter value of $\tau_e$ $\sim$ 250 Myr. This means that tidal circularization is expected to happen in a few hundreds of Myr. 
Similarly, we computed the tidal decay timescale $\tau_a$, using equation (2) in Jackson et al. We obtained $\tau_a$ $\sim$ 10 Gyr, which is longer than the age of the system.
By comparing $\tau_a$ with $\tau_e$, it is clear that circularization is expected to occur well before tidal engulfment, thus we should expect to find two different populations
of short-period planets around first ascending RGB stars, those in which the host star is close to the base of the RGB, thus they have retained their primordial eccentricity, and those that are located around more evolved stars, that are expected to have nearly circular orbits. The discovery of new planets
like that presented here will allow us to confirm this prediction, while at the same time we can use them to calibrate the tidal efficiencies in fully convective RGB stars. Unfortunately, detecting transiting planets around more evolved stars whose radii are significantly larger, is still very challenging due to the reduced transit depth and longer duration of the transit.

\subsection{Summary}

In this paper we present the discovery of a 9.2-day orbit transiting planet around the giant star EPIC\,228754001, from the high precision photometric data taken by the $K2$ mission.
A further spectroscopic follow-up allowed us to confirm the planetary nature of the periodic transit observed in the $K2$ data. Our discovery was made independently from Grunblatt et al (\cite{GRU17}), who announced the result as we were writing up this paper.
Based on the combined photometric and RV analysis, which includes all radial velocities available including the Keck/HIRES RVs of Grunblatt et~al. (2017), we derive a planetary mass of 0.50 $\pm$ 0.01\mjup and an eccentricity of 0.29 $\pm$ 0.05, making EPIC\,228754001\,$b$ the most eccentric planetary companion among all known short-period ($P \lesssim$ 50 days) planet orbiting giant stars. Using the high precision photometric data, we performed an asteroseismic analysis, from which we derived a stellar mass and radius of 1.19 $\pm$ 0.04 and 4.11 $\pm$ 0.05, respectively. \newline
In addition, we put this planet into context, by comparing its orbital properties to those systems that have been found around giant stars. We concluded that the existence of transiting systems like EPIC\,228754001\,$b$ provide an observational support for a different formation and migration scenario for planets in more massive protoplanetary disks around more massive stars. \newline
Finally, we discussed about the orbital evolution (circularization and tidal decay timescales) for this system. From this analysis, we concluded that more eccentric systems like this one might be found by transit surveys around giant stars close to the base of the RGB, while a population of planet in nearly circular orbits is expected to be found around stars that are slightly more evolved than EPIC\,228754001.

\begin{acknowledgements}
R.B., N.E. and A.J. acknowledge  support from the Ministry for the Economy, Development and Tourism Programa Iniciativa Cient\'ifica Milenio through grant IC 120009, awarded to the Millenium Institute of Astrophysics.  A.J. acknowledges support by Fondecyt grant 1171208 and partial support by CATA-Basal (PB06, CONICYT). N.E. acknowledges support from Financiamiento Basal PFB06. 
\end{acknowledgements}

\end{document}